\documentclass[aip,jcp,preprint,amssymb]{revtex4-1}

\usepackage{graphicx}
\usepackage{color}
\usepackage{bm}
\usepackage{amsmath}
\usepackage{multirow}
\usepackage{booktabs, dcolumn}

\begin{document}

\title{Gas-phase structural isomer identification by Coulomb explosion of aligned molecules}

\author{Michael Burt}
\thanks{These authors contributed equally to this work.}
\author{Kasra Amini}
\thanks{These authors contributed equally to this work.}
\affiliation{The Chemistry Research Laboratory, Department of Chemistry, University of Oxford, 12 Mansfield Road, Oxford OX1 3TA, United Kingdom}
\author{Jason W. L. Lee}
\affiliation{The Chemistry Research Laboratory, Department of Chemistry, University of Oxford, 12 Mansfield Road, Oxford OX1 3TA, United Kingdom}
\author{Lars Christiansen}
\affiliation{Department of Chemistry, Aarhus University, Langelandsgade 140, DK-8000 Aarhus C, Denmark}
\author{Rasmus R. Johansen}
\affiliation{Department of Physics and Astronomy, Aarhus University, Ny Munkegade 120, DK-8000 Aarhus C, Denmark}
\author{Yuki Kobayashi}
\affiliation{Department of Chemistry, University of California, Berkeley, CA 94720, United States of America}
\author{James D. Pickering}
\affiliation{Department of Chemistry, Aarhus University, Langelandsgade 140, DK-8000 Aarhus C, Denmark}
\author{Claire Vallance}
\affiliation{The Chemistry Research Laboratory, Department of Chemistry, University of Oxford, 12 Mansfield Road, Oxford OX1 3TA, United Kingdom}
\author{Mark Brouard} \email{mark.brouard@chem.ox.ac.uk}
\affiliation{The Chemistry Research Laboratory, Department of Chemistry, University of Oxford, 12 Mansfield Road, Oxford OX1 3TA, United Kingdom}
\author{Henrik Stapelfeldt} \email{henriks@chem.au.dk}
\affiliation{Department of Chemistry, Aarhus University, Langelandsgade 140, DK-8000 Aarhus C, Denmark}
\date{24 January 2018}


\begin{abstract}
The gas-phase structures of four difluoroiodobenzene and two dihydroxybromobenzene isomers were identified by correlating the emission angles of atomic fragment ions created following femtosecond laser-induced Coulomb explosion. The structural determinations were facilitated by confining the most polarizable axis of each molecule to the detection plane prior to the Coulomb explosion event using one-dimensional laser-induced adiabatic alignment. For a molecular target consisting of two difluoroiodobenzene isomers, each constituent structure could additionally be singled out and distinguished.
\end{abstract}


\maketitle


Laser-induced molecular Coulomb explosion is the process whereby an intense femtosecond laser pulse detaches several electrons from a molecule, breaking it into cationic fragments. If the axial recoil approximation is satisfied, the created fragment ions recoil along the original bond axes of their parent molecule, allowing laser-induced Coulomb explosions to be used for two purposes: the first concerns how molecules are oriented in space at the instant the laser pulse arrives, and is measured by determining the emission directions of the fragment ions with respect to one or more fixed axes. This method has been used as a probe for a large number of laser-induced alignment and orientation experiments.\cite{larsenaligning1999, dooley2003direct, stapelfeldtcolloquium2003, pentlehner2013impulsive} The second concerns molecular structures, which can be identified by correlating the measured fragment momenta. These latter experiments have revealed static molecular geometries and, with femtosecond pump-probe schemes, structural and reaction dynamics. \cite{Stapelfeldt1995, Kitamura2001, Ergler2005a, Legare2006b, madsen2009manipulating, Pitzer2013, Christensen2014, Ibrahim2014, Erk2014, Christensen2015, Ablikim2016, Pitzer2016c, Amini2017a, Burt2017, Savelyev2017}

Coulomb explosion research on diatomic and triatomic molecules has demonstrated that covariance and coincidence analyses are efficient and powerful approaches for identifying the correlations between the emission directions of different fragment ions. \cite{Frasinski1989,  Bryan2000, Hasegawa2001, Legare2005, Gagnon2008a, Frasinski2016} This is even more evident in studies of larger and more complex molecules, towards which interest has turned over the past decade.\cite{Hansen2012, Pitzer2013, Slater2014, Pitzer2016c} Covariance analysis of Coulomb explosion fragments has also been applied to molecules that were pre-aligned by adiabatic laser pulses.\cite{Hansen2012, Slater2014, Slater2015} In these experiments, this alignment placed the molecules in well-defined spatial orientations with respect to the imaging detector, and thereby increased the structural information obtained from the recorded fragment momenta. The purpose of this work is to further develop Coulomb explosion imaging of pre-aligned targets as a method to determine the structures of polyatomic molecules. In particular, we demonstrate that four difluoroiodobenzene (DFIB) isomers can be unequivocally distinguished by analyzing the angular correlations of specific fragment ions. Furthermore, we show that for a mixture of two DFIB isomers the constituent structures can be singled out and identified. Finally, to demonstrate that our method is also capable of identifying isomers when only one substituent is a halogen atom, experiments were carried out on two isomers of dihydroxybromobenzene (DHBB).

In the following experiments a pulsed molecular beam of approximately 1 mbar DFIB or DHBB in 80 bar He was expanded into a velocity-map imaging mass spectrometer and crossed perpendicularly by two pulsed laser beams.\cite{Chandler1987, Eppink1997} The pulses in the first beam (alignment pulses: duration, $\tau$ = 10\,ns (FWHM); $\lambda$ = 1064\,nm; I$_0$ = 5.0 $\times$ 10$^{11}$\,W\,cm$^{-2}$) were used for adiabatic alignment, and the pulses in the second beam (probe pulses: $\tau$ = 35\,fs (FWHM); $\lambda_\text{center}$ = 800\,nm; I$_0$ = 3.0 $\times$ 10$^{14}$\,W\,cm$^{-2}$) were used to Coulomb explode the molecules. Each probe pulse was synchronized to the peak of an alignment pulse to ensure that only the most strongly aligned molecules were probed. The ions produced by the probe pulse were then extracted onto a microchannel plate detector backed by a phosphor screen, where their positions ($x$, $y$) and arrival times ($t$), relative to an external trigger, were recorded at 10\,Hz to a precision of 40\,ns with the $324 \times 324$ pixel PImMS2 sensor.\cite{Clark2012, John2012} Further details of the experimental setup are described elsewhere. \cite{Hansen2011,Slater2014,Slater2015}

\begin{figure}
\centering
\includegraphics{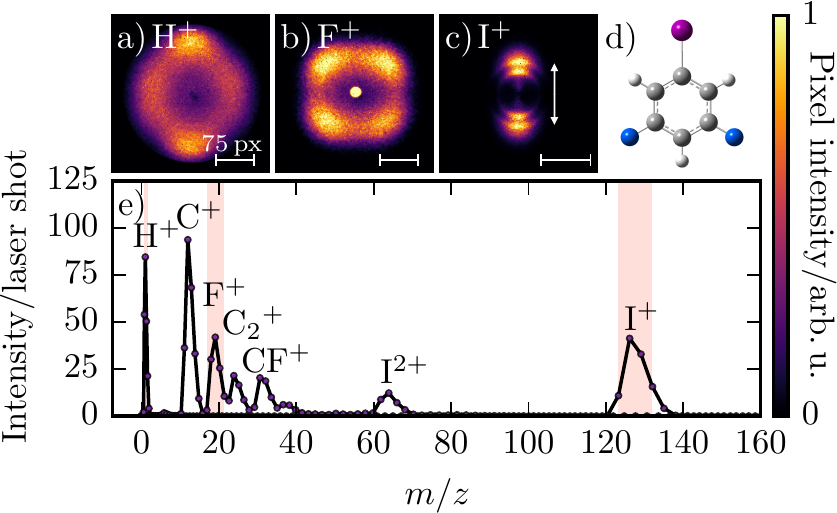}
\caption{(a)-(c): The velocity-map images of H$^{+}$, F$^{+}$, and I$^{+}$, and (e): the corresponding mass spectrum resulting from the Coulomb explosion of one-dimensionally aligned 3,5-difluoroiodobenzene by the probe pulse. Both the alignment and the probe pulse are linearly polarized in the direction shown by the arrow on panel (c). (d): Sketch of the molecular structure of 3,5-DFIB.
\label{fig1}}
\end{figure}

Figure\ \ref{fig1} demonstrates the observables acquired from the Coulomb explosion of 3,5-DFIB. The two-dimensional velocity-map images of the I$^{+}$, F$^{+}$ and H$^{+}$ fragments of 3,5-DFIB are presented in panels (a)-(c), and its structure is provided in panel (d). The images were recorded with both the alignment and the probe pulse present. Each ion image represents data accumulated over 50,000 laser shots and was obtained by adding all ion hits with arrival times corresponding to the relevant mass-to-charge ratios ($m$/$z$), which are indicated by the red shaded areas in panel (e). The pronounced angular confinement of the I$^{+}$ fragments along the central vertical axis (Fig.\ \ref{fig1}(c)) demonstrates that the 3,5-DFIB molecules are confined along their most polarizable axis, the C-I bond, by the polarization vector of the alignment pulse. This is the expected result and matches previous work, from which it is also known that the plane of the molecule is free to rotate around the C-I-axis. \cite{viftrup2007holding} Note that any other DFIB isomer will be aligned in the same manner because the polarizability tensor is essentially independent of the location of the F atoms with respect to the I atom.

\begin{figure}
\centering
\includegraphics{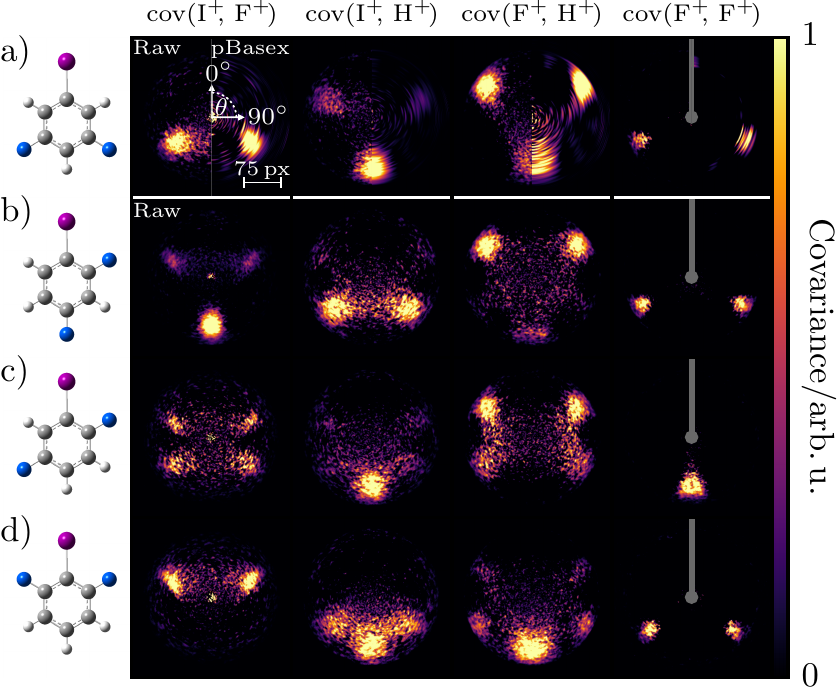}
\caption{Recoil-frame covariance images of (a) 3,5-; (b) 2,4-; (c) 2,5-; and (d) 2,6-difluoroiodobenzene. Each image is labeled as cov(X$^{+}$, Y$^{+}$) for a reference ion X$^{+}$ and partner ion Y$^{+}$, and represents the velocity distribution of Y$^{+}$ with respect to the velocity of X$^{+}$,  here set to point towards 0$^{\circ}$. The right halves of the images in (a) illustrate the result of Abel inverting the raw data using \textsc{pbasex},\cite{Garcia2004} whereas the left halves show the raw ion images. The intensities of each image are normalized to their most intense pixels, and only positive covariances are shown. The autovariance in the cov(F$^{+}$, F$^{+}$) images has been masked for clarity. The respective molecular reference frames of each isomer are shown in the leftmost column; C: gray, H: white, F: blue, I: purple.
\label{fig2}}
\end{figure}

Although the velocity-map images of the 3,5-DFIB fragments provide clues about the corresponding molecular structure, covariance analysis of the correlations between the fragment momenta is required to distinguish the substituent positions and bond angles of each isomer. For each of the four DFIB isomers considered here, four recoil-frame covariance images illustrating the relative velocities of a pair of recoiling ions are shown. These are collected in Fig.\ \ref{fig2}, where each covariance image, cov(X$^{+}$, Y$^{+}$), represents the recoil velocity distribution of a partner ion Y$^{+}$ with respect to the recoil direction of a reference ion X$^{+}$.\cite{Slater2014, Slater2015} To understand the potential for structural identification offered by the covariance images, assume that the only knowledge available from the mass spectra of the isomers is that they are each benzene molecules containing at least one F and one I atom. For all four cases studied the cov(I$^{+}$, I$^{+}$) images (not displayed) only contain autovariance features, indicating that each molecule has only one I atom. Starting with row (a) the cov(I$^{+}$, F$^{+}$) image shows that F$^{+}$ ions are ejected at $\sim$$\pm$120$^{\circ}$ with respect to I$^{+}$. This implies that the molecule contains one or two F atoms in the 3 or 5 positions. The cov(F$^{+}$, F$^{+}$) image shows that emission of one F$^{+}$ is correlated with another F$^{+}$ departing $\sim$120$^{\circ}$ away (i.e. the molecule must contain two F atoms placed in the 3 and 5 positions). We therefore conclude that the molecule is unambiguously the 3,5-DFIB isomer.  This is corroborated by the cov(I$^{+}$, H$^{+}$) image showing emission of H$^{+}$ at 180$^{\circ}$ and $\sim$60$^{\circ}$ (i.e. H atoms in the 2, 4 and 6 positions), and also by the cov(F$^{+}$, H$^{+}$) image showing that the H atoms are placed para or ortho to the F atoms.

Turning to row (b) the cov(I$^{+}$, F$^{+}$) image shows that this molecule has an F atom in position 4 and at least one in position 2 and/or 6. The cov(I$^{+}$, H$^{+}$) image shows that there are H atoms in positions 3 and 5 and in 2 or 6. Together these two covariance images allow us to conclude unequivocally that the isomer is 2,4-DFIB. In row (c) the cov(I$^{+}$, F$^{+}$) image shows that there is at least one F atom in position 2 and/or 6 and at least one F atom in position 3 and/or 5. The cov(F$^{+}$, F$^{+}$) image shows that the F atoms are placed opposite (180$^{\circ}$) to each other (i.e. there are only two F atoms in the molecule). We conclude, again unequivocally, that the molecule is the 2,5-DFIB isomer. Finally, in row (d) the cov(I$^{+}$, F$^{+}$) image shows that the molecule contains one or two F atoms in position 2 and/or 6. The cov(F$^{+}$, F$^{+}$) image shows that there are two F atoms, recoiling at $\sim$120$^{\circ}$. We again conclude unambiguously that the molecule is the 2,6-DFIB isomer.

A more careful analysis of the relative ion recoil angles is obtained by applying an inverse Abel transform to each covariance image with \textsc{pbasex} (see Fig.\ \ref{fig2}(a)) and reproducing the resulting angular distributions with Gaussian functions. \cite{Garcia2004, Slater2014} The results are given in Table 1 with their standard deviations. It appears that the relative angles do not all faithfully represent the angles between the bonds in the parent molecule. This is to be expected because the final emission directions of the fragment ions are influenced to some degree by their mutual electrostatic repulsion as they fly away from the molecule. A more precise determination of the parent molecule structure therefore requires that the effect of the fragment-fragment repulsion be understood. This is possible by simulating the Coulomb explosion dynamics with classical ion trajectory calculations, but is not needed for the current work. \cite{Slater2015}

 \begin{table}
 \caption{\label{tables1}Difluoroiodobenzene fragment recoil angles.}
 \begin{ruledtabular}
 \begin{tabular}{c D{,}{\pm}{1.2} D{,}{\pm}{1.2} D{,}{\pm}{1.2} D{,}{\pm}{1.2} }
Isomer & \multicolumn{1}{c}{ $\measuredangle$(I$^{+}_{,}$F$^{+}$)$^{\circ}$} &  \multicolumn{1}{c}{$\measuredangle$(I$^{+}_{,}$H$^{+}$)$^{\circ}$} &  \multicolumn{1}{c}{$\measuredangle$(F$^{+}_{,}$H$^{+}$)$^{\circ}$} &  \multicolumn{1}{c}{$\measuredangle$(F$^{+}_{,}$F$^{+}$)$^{\circ}$} \\
\hline
\multirow{2}{*}{3,5} & 121\,,\,7 & 74\,,\,8 & 60\,,\,7 & 113\,,\,6  \\
 & & 180\,,\,12 & 180\,,\,25 & \\
\hline
\multirow{3}{*}{2,4} & 72\,,\,7 & 63\,,\,9 & 58\,,\,7 & 117\,,\,6 \\
 & 180\,,\,11 & 127\,,\,8 & 124\,,\,5 & \\
 & & & 180\,,\,21 & \\
\hline
\multirow{4}{*}{2,5} & 71\,,\,7 & 73\,,\,8 & 62\,,\,8 & 180\,,\,11 \\
 & 122\,,\,8 & 124\,,\,8 & 112\,,\,10 & \\
 & & 180\,,\,12 & & \\
\hline
\multirow{3}{*}{2,6} & 71\,,\,8 & 123\,,\,8 & 72\,,\,15 & 125\,,\,6 \\
 & & 180\,,\,14 & 116\,,\,7 & \\
 & & & 180\,,\,18 & \\
 \end{tabular}
 \end{ruledtabular}
 \label{tab:DFIB}
 \end{table}

\begin{figure}
\centering
\includegraphics{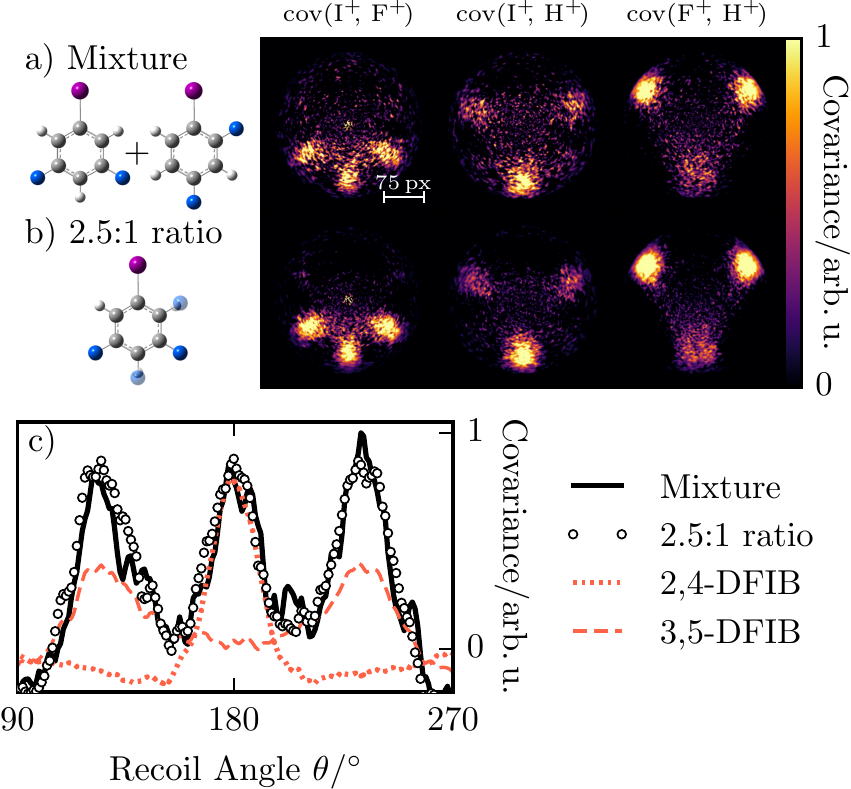}
\caption{Recoil-frame covariance images of (a) a mixture of difluoroiodobenzene isomers and (b) the 3,5- and 2,4-isomer data in Fig.\ \ref{fig2} summed in a 2.5:1 ratio. The labeling and representation style is the same as that used in Fig.\ \ref{fig2}. (c) The angular distributions of the F$^{+}$ ions from the cov(I$^{+}$, F$^{+}$) images for the mixed sample compared with the constituent 3,5- and 2,4-DFIB isomers summed in a 2.5:1 ratio (see text). The pictured isomer structures in (a) are added and superimposed in (b) to illustrate the overlapping molecular reference frames.
\label{fig3}}
\end{figure}

Next we demonstrate that the recoil-frame covariance method is also capable of distinguishing structural isomers within a sample mixture. To do so we recorded ion images for a sample consisting of two DFIB isomers in an unknown ratio. The results are displayed in row (a) of  Fig.\ \ref{fig3}. The cov(I$^{+}$, F$^{+}$) image shows that the F$^{+}$ ions recoil at approximately $\pm$120$^{\circ}$ and 180$^{\circ}$ with respect to the direction of the I$^{+}$ ion. The only isomer that gives a signal at 180$^{\circ}$ is 2,4-DFIB so that one must be present in the sample. The signal at $\pm$120$^{\circ}$ may originate from either 3,5-DFIB or 2,5-DFIB. However, the cov(I$^{+}$, H$^{+}$) image resembles that of 3,5-DFIB much more than that of 2,5-DFIB (see Fig.\ \ref{fig2}), strongly indicating that 3,5-DFIB is the other component of the unknown sample. To quantify this assignment we determined the angular distributions of the reference ions from the covariance images in row (a) and tried to match them by a linear combination of the angular distribution of the reference ions from the 2,4-DFIB and the 3,5-DFIB covariance images in Fig.\ \ref{fig1}. The ratio of the two isomer contributions is a free parameter.  Figure\ \ref{fig3}(c) shows the result for the F$^{+}$ angular distribution from the cov(I$^{+}$, F$^{+}$) images. The agreement between the mixed sample and the result of adding the 3,5-DFIB and 2,4-DFIB isomers in a 2.5:1 ratio is excellent. Similar agreement is found for the angular distribution of H$^{+}$ ions from the cov(I$^{+}$, H$^{+}$) and the cov(F$^{+}$, H$^{+}$) images. The images in row (b), obtained by adding the covariance images from the individual 3,5-DFIB to 2,4-DFIB isomers in a ratio of 2.5:1, also agree very well with the covariance images from the unknown sample. As a check we also tried to reproduce the covariance images and derived angular distributions in Figs.\ \ref{fig3}(a) and (c) by adding the individual 2,5-DFIB and 2,4-DFIB covariance images. No ratio of the isomers could produce good agreement with the mixture.

\begin{figure}
\includegraphics{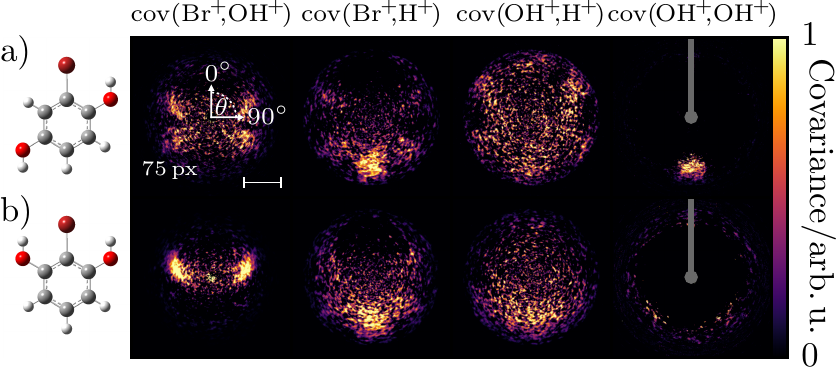}
\caption{Recoil-frame covariance images of (a) 2,5- and (b) 2,6-dihydroxybromobenzene using the same labeling and representation style as described in the caption to Fig.\ \ref{fig2}. The respective molecular reference frames of each isomer are shown in the leftmost column; C: gray, H: white, O: red, Br: brown.
\label{fig4}}
\end{figure}

Finally, we carried out experiments on 2,5- and 2,6-dihydroxybromobenzene. Similar to the DFIB studies, we assume that we only know that the sample molecules are benzenes substituted with Br and OH and show that the covariance images in Fig.\ \ref{fig4} allow the structural isomer contained in each sample to be identified.  In row (a) the cov(Br$^{+}$, OH$^{+}$) image has the same four-peak structure as the cov(I$^{+}$, F$^{+}$) image for the 2,5-DFIB isomer, and the cov(OH$^{+}$, OH$^{+}$) image has one feature at 180$^{\circ}$, similar to the cov(F$^{+}$, F$^{+}$) image of 2,5-DFIB. Applying the same arguments made for DFIB, we conclude that the data in row (a) arises from the 2,5-DHBB isomer. Likewise, the upper two-peak structure of the cov(Br$^{+}$, OH$^{+}$) image in row (b) closely resembles that of the cov(I$^{+}$, F$^{+}$) image for 2,6-DFIB, and the lower two-peak structure in the cov(OH$^{+}$, OH$^{+}$) image is similar to the structure in the cov(F$^{+}$, F$^{+}$) image for 2,6-DFIB. Again, using the same argumentation as for DFIB we unambiguously conclude that the data in row (b) stems from 2,6-DHBB.

In the DHBB measurements O$^{+}$ and OH$^{+}$ were not resolved in the mass spectrum. The OH$^{+}$ images therefore contain some fraction of O$^{+}$. The emission directions of the latter may deviate from the initial C-O bond direction in the parent molecules due to the effect of the H or H$^{+}$ that necessarily leaves from the OH group when O$^{+}$ is formed. This could be the reason why the cov(Br$^{+}$, OH$^{+}$) and cov(OH$^{+}$, OH$^{+}$) images are less sharp than the corresponding cov(I$^{+}$, F$^{+}$) and cov(F$^{+}$, F$^{+}$) images of DFIB. It may also be part of the reason why the average recoil angle between the Br$^{+}$ and the OH$^{+}$/O$^{+}$ from the ortho OH substituent is larger (78$^{\circ}$/81$^{\circ}$ for 2,5- and 2,6-DHBB, see Table 2) than the average recoil angle between I$^{+}$ and the F$^{+}$ from the ortho F atom (71$^{\circ}$, see Table 1) in the corresponding DFIB isomers. Turning to the covariance images involving H$^{+}$ it appears that the cov(Br$^{+}$, H$^{+}$) and cov(OH$^{+}$, H$^{+}$) images (middle panels in row (a) and (b) of  Fig.\ \ref{fig4}) exhibit broader recoil angles than the analagous cov(I$^{+}$, H$^{+}$) and cov(F$^{+}$, H$^{+}$) images in Fig.\ \ref{fig2}. The reason is that for DHBB the H$^{+}$ can arise not only from the H atoms bonded directly to the benzene ring but also from the H atoms on the two hydroxyl groups.  The latter signal depends on the rotation of the O-H bonds around their connecting C-O bonds, which for the ortho hydroxyl substituents is likely dictated by the degree of hydrogen bonding with bromine. With greater angular resolution the recoil-frame covariance images involving H$^{+}$ could in principle provide more detailed information about the conformers of the DHBB isomer. 
\begin{table}
 \caption{\label{tables2}Dihydroxybromobenzene fragment recoil angles.}
\begin{ruledtabular}
\begin{tabular}{c D{,}{\pm}{1.2} D{,}{\pm}{1.2} D{,}{\pm}{1.2} D{,}{\pm}{1.2} }
Isomer & \multicolumn{1}{c}{$\measuredangle$(Br$^{+}_{,}$OH$^{+}$)$^{\circ}$} &  \multicolumn{1}{c}{$\measuredangle$(Br$^{+}_{,}$H$^{+}$)$^{\circ}$} &  \multicolumn{1}{c}{$\measuredangle$(OH$^{+}_{,}$H$^{+}$)$^{\circ}$} &  \multicolumn{1}{c}{$\measuredangle$(OH$^{+}_{,}$OH$^{+}$)$^{\circ}$} \\
\hline
\multirow{4}{*}{2,5} & 78\,,\,8 & 72\,,\,17 & 0\,,\,15 & 180\,,\,11 \\
 & 121\,,\,8 & 128\,,\,7 & 65\,,\,11 & \\
 & & 180\,,\,12 & 114\,,\,10 & \\
 & & & 180\,,\,21 & \\
 \hline
\multirow{3}{*}{2,6} & 81\,,\,9 & 74\,,\,12 & 0\,,\,17 & 133\,,\,9 \\
 & & 129\,,\,11 & 180\,,\,45 & \\
 & & 180\,,\,14 & & \\
\end{tabular}
\end{ruledtabular}
\label{tab:DHBB}
\end{table}

In conclusion, we demonstrated that four isomers of DFIB could be unambiguously identified by correlating the fragment ion momenta created following femtosecond laser-induced Coulomb explosion of one-dimensionally aligned molecules. The isomer-specific recoil-frame covariance images were then used as references for identifying two isomers, including their ratio, in a mixed sample. The mixture resembles a situation often encountered in femtosecond time-resolved photochemistry, where a small fraction of molecules are excited before undergoing isomerization. As such, our results point towards the use of laser-induced Coulomb explosion as a tool to image the evolving structure of excited molecules within a large background of unexcited (ground state) molecules. Finally, our methods were employed to identify two isomers of DHBB as an illustration that fragment ions other than halogens, here OH$^{+}$, also enable structural determinations. This indicates the potential application of Coulomb explosion imaging to a much broader class of molecules than have previously been studied. Along these lines, we additionally note that the current work was based on two-fold covariances, but three-fold covariance analysis could further improve and extend the structural identification capabilities provided sufficient data statistics are available. \cite{Pickering2016}


\begin{acknowledgments}
The M.B.\ and C.V.\ groups gratefully acknowledge support from the UK EPSRC (Programme Grant No.\ EP/L005913/1), the EU (FP7 ITN 'ICONIC', Project No.\ 238671), and the STFC (PNPAS award and  mini-IPS Grant No.\ ST/J002895/1). Y.K. is grateful to the Army Research Office (Grant No. W911NF-14-1-0383). H.S. acknowledges support from the European Research Council-AdG (Project No. 320459, DropletControl) and from the EU via the MEDEA project within the Horizon 2020 research and innovation programme under the Marie Sk\l{}odowska-Curie Grant Agreement No. 641789.
\end{acknowledgments}


\bibliography{isomer-covariance}


\end{document}